\pdfoutput=1 
\documentclass{JINST}

\title{Precise On-line Position Measurement for Particle Therapy}

\author{O. Actis\thanks{Corresponding author.}~,
D. Meer and S. K{\"o}nig\\
\llap{$^a$}Paul Scherrer Institute (PSI),\\
 5232 Villigen PSI, Switzerland\\

E-mail: \email{oxana.actis@psi.ch}}

\abstract{An on-line beam position monitoring and regular beam stability tests are of utmost importance for the Quality Assurance (QA) of the patient treatment at any particle therapy facility. The Gantry$\hspace{0.5 mm}2$ at the Paul Scherrer Institute uses a strip ionization chamber for the on-line beam position verification. The design of the strip chamber placed in the beam in front of the patient allows for a small beam penumbra in order to achieve a high-quality lateral beam delivery. 
The position error of 1 mm in a lateral plane (plane perpendicular to the beam direction) can result in a dose inhomogeneity of more than $5 \%$. Therefore the goal of Gantry 2 commissioning was to reach a sub-millimeter level of the reconstruction accuracy in order to bring a dose uncertainty to a level of $1 \%$. In fact, we observed that for beams offered by Gantry 2 signal profiles in a lateral plane can be reconstructed with a precision of 0.1 mm. This is a necessary criterion to perform a reliable patient treatment. The front end electronics and the whole data processing sequence have been optimized for minimizing the dead time in between two consecutive spots to about 2 ms: the charge collection is performed in about 1 ms, read-out takes place in about 100 $\mu$s while data verification and logging are completed in less than 1 ms. }


\begin{document}

\section{Introduction}\label{sec:intro}
Particle therapy has proven to be very effective for cancer treatment all over the world. Currently around 40 proton and carbon-ion therapy centers are active and patients are treated worldwide; more centers will be constructed in the next future. In particle therapy the properties of the Bragg peak are exploited, and it is possible to deliver a high dose to the tumor being treated while minimizing effects on healthy tissues. In this way, the risk of post-treatment complications is reduced with respect to conventional radio therapy \cite{bib0,bib00}.

The Gantry 2 at the Paul Scherrer Institute (PSI) is a therapy delivery system which operates using proton beams of energies ranging from 70 MeV to 230 MeV \cite{bib1}. The Gantry$\hspace{0.5 mm}2$ employs an active spot-scanning technique which allows to follow the tumor shape by means of a fast change of the energy beam for scanning in depth. In addition, two sweeper magnets deflect the proton beam to the required position in a plane perpendicular the beam direction (lateral plane). The Gantry$\hspace{0.5 mm}2$ offers beams with a dose per spot which spans from $10^5$ to $10^8$ protons per spot. The beam size can vary from 1.1 mm to 2.3 mm in the Gantry nozzle. The precision of the dose delivery in depth depends on an accurate energy selection and a reliable description of the propagation medium. Since the homogeneity of the delivered dose distribution directly depends on the lateral position accuracy, the signal of all provided parameters must be reconstructed with a sub-millimeter precision. The on-line dose and position monitoring of the proton beam during the treatment as well as regular stability checks are crucial for the Quality Assurance (QA) in order to ensure a reliable high-quality patient treatment.

\section{Materials and Methods}\label{sec:m_and_m}
There are many different methods which can be used for the lateral position verification with the highest position resolution at a single-particle count level. However, in the case of on-line position verification for particle therapy, the detector has to be placed right in the beam and, as a consequence, should have possibly low material budget in order to reduce the multiple scattering during the beam delivery. This criterion is of critical importance for protons since the effect of Multiple Coulomb Scattering is much higher for these particles than for carbon ions.

\begin{figure}[tp] 
\centering
\includegraphics[width=.65\textwidth]{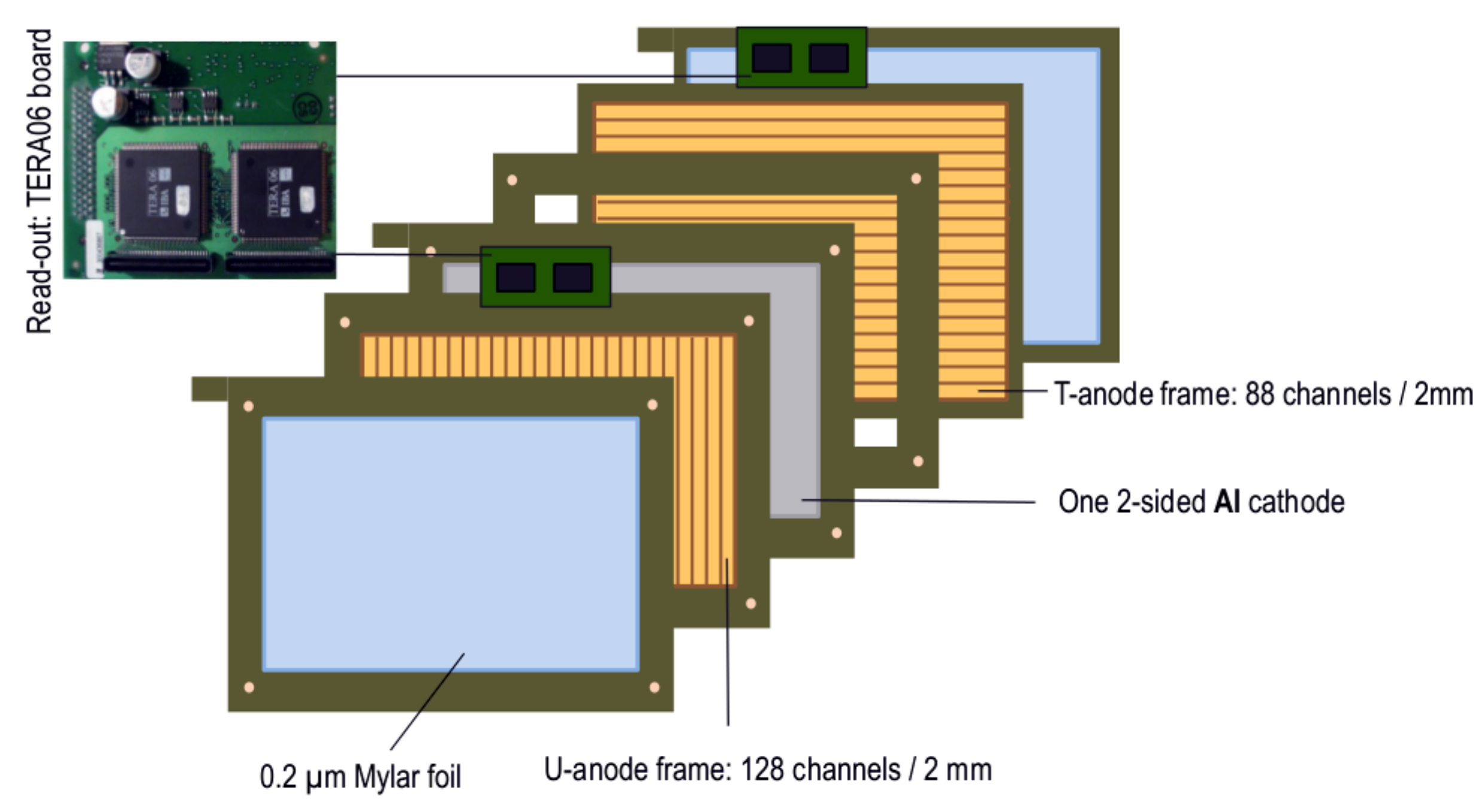}
\caption{Schematic view of the ionization strip chamber.}
\label{fig:01}
\end{figure}

For the on-line lateral position verification Gantry 2 uses an ionization strip chamber. This device minimizes beam disturbance and showed extremely stable operation over almost 20 years at previous therapy system, Gantry 1 \cite{bib2}. This chamber covers the full scanning area of 20 by 12 cm with two perpendicular planes of 88 and 128 strips and a strip size of 2 mm. The chamber has one cathode plate in the middle which is a 20 ${\mu}$m Mylar foil 2-side coated with 20 nm of aluminum. Anodes are 50 ${\mu}$m-thick Kapton foils with 17 ${\mu}$m copper electrodes. The entrance and exit windows of the detector are 0.2 ${\mu}$m stretched double-side metalized Mylar foils, as that used for cathode. The counting gas of the chamber is an ambient air and the gap between cathode and anode plates is adjustable by detector design. We use an air gap of 1 cm and a high voltage of 1800 V. The fixed amount of subtracted charge in a current-to-frequency converter (quantum of charge) is set to 200 fC; however, the electronics design gives the possibility to choose between 100 and 800 fC. The schematic view of the strip chamber is shown in Figure \ref{fig:01}. This detector was developed by a collaboration of the TERA foundation, the INFN of Torino and PSI, and is currently commercialized by the DE.TEC.TOR company \cite{bibDETECTOR}.

\begin{figure}[tp] 
\centering
\includegraphics[width=.8\textwidth]{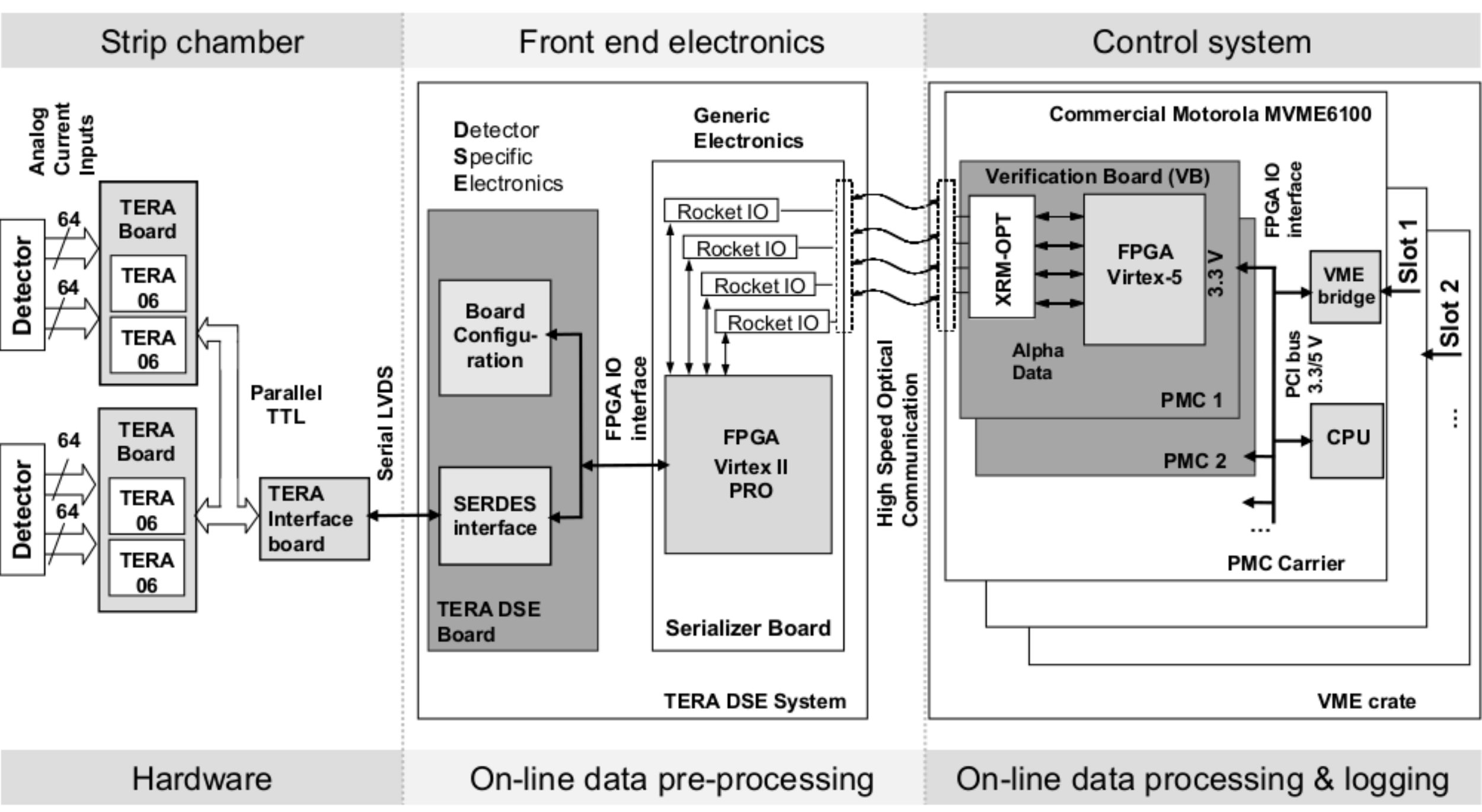}
\caption{Ionization strip chamber read-out and data processing sequence. First charge collected from the detector is digitized by ADC (TERA Board). Afterwards data is serialized (serdes interface) and is passed to an FPGA for pre-processing. Finally the data is sent for further processing and verification within the therapy control system.}
\label{fig:02}
\end{figure}

Figure {\ref{fig:02}} illustrates the data processing sequence together with a block diagram of the read-out electronics. The principle of the strip chamber read-out is based on a current-to-frequency converter followed by a counter: a TERA 06 board with two 64-channel Very-Large Scale Integration (VLSI) chips \cite{bibTERA}. Digital signal from both anode plates are sent to the TERA Detector Specific Electronics (DSE) system which consists of two parts: the TERA DSE board and the Serializer Board. A Field-Programmable Gate Array (FPGA) is installed on a second board, where a firmware pre-calculation of the beam position is performed and sent via a high speed optical link to the control system. The final calculation of the position and size of the spot is done through a software: 
the spot position is propagated to the iso-center taking into account the Gantry 2 beam optics and cross-checked with the expected value. Afterwards these data are logged into a binary file together with other dose field parameters within less than one millisecond time. 

In addition to the on-line position monitor, an ionization strip chamber of the same type is used for regular position cross-checks. Moreover, two smaller strip chambers with an active area of 7 by 7 cm and a strip size of 2.2 mm are used for the daily verification of the beam size, position and direction. 

\section{Results}\label{sec:res}
\begin{figure}[tp] 
\centering
\includegraphics[width=0.85\textwidth]{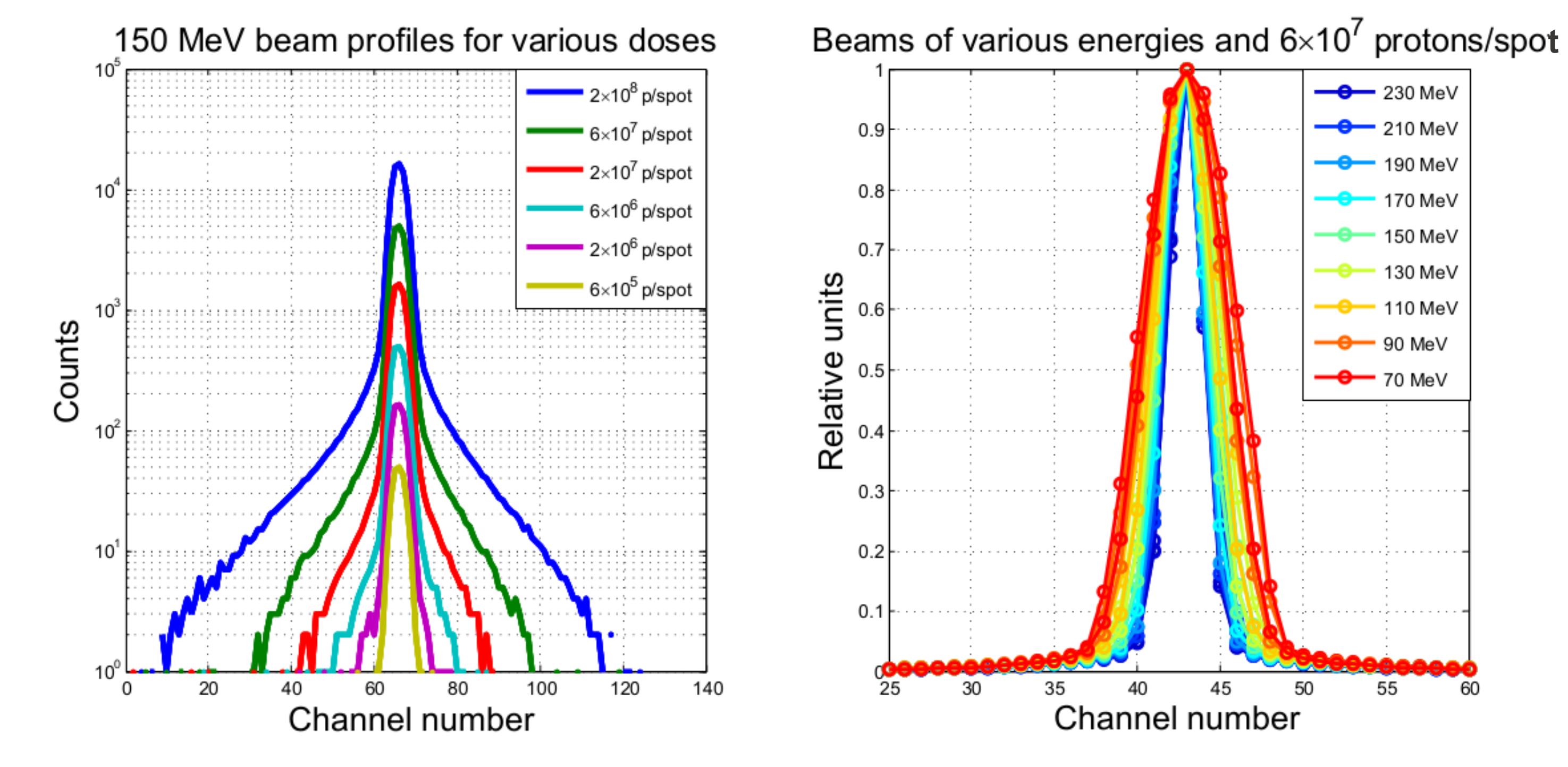}
\caption{Reconstructed beam profiles for the nozzle strip chamber. Left: 150 MeV beams applied with weights varying from $6\times10^5$ to $2\times10^8$ protons per spot. Right: beams of $6\times10^7$ protons per spot and energies varying from 70 MeV to 230 MeV in 20 MeV steps. }
\label{fig:03}
\end{figure}

A position deviation of more than one millimeter can lead to a dose fluctuation of several percents; therefore the required position precision has to reach the sub-millimeter level. The Gantry$\hspace{0.5 mm}2$ strip monitor allows an on-line position and shape control of the full range of beams available at our machine with a sub-millimeter precision. Signals varying from high-weighted spots down to the lowest dose used by our therapy planning system (order of tenth of a milligray) can be reconstructed with the required precision level due to the low detector noise. Figure \ref{fig:03} shows typical reconstructed Gantry 2 beam profiles. The left plot shows reconstructed signals produced by 150 MeV beams with dose varying from $6\times{10^5}$ to $2\times{10^8}$ protons per spot. The minimal dose corresponds to the lowest dose spot used in therapy planning. In this figure one can clearly see that also for the lowest weighted spot the signal-to-noise ratio is still at a very good level. The plot on the right-hand side demonstrates reconstructed profiles for high-weighted signals ($6\times{10^7}$ protons per spot) of energies ranging from 70 to 230 MeV in 20 MeV steps. Even for high energies, where the signal peak is very narrow, the detector granularity of 2 mm is more than sufficient for a high-precision signal reconstruction. The overall precision for the lateral position reconstruction is at a level of 0.1 mm for the whole range of available signals.

\section{Conclusions and Outlook}\label{sec:conc}
Strip ionization chambers have proven to be an appropriate verification and QA tool for the scanning proton beam therapy system. A suitable design has allowed to operate in a simple, efficient and  stable way over several years. The system demonstrates a sub-millimeter precision for the position reconstruction which is needed to ensure dose homogeneity of 1\% level. This precision guarantees a patient treatment quality at the highest achievable level. The data acquisition and processing are optimized in a way that the total read-out time remains below 100 $\mu$s. Therefore the total dead time between two spots is dominated by the physical process of charge collection in the ionization chamber.

However, the sector of beam delivery technologies is developing and the design of the position detector could require further improvements. This can be done using a lower material budget, optimizing the air gap, the strip size and the front end electronics as well as using a different gas instead of air to increase charge mobility.

\end{document}